\documentclass[journal]{IEEEtran}
\ifCLASSINFOpdf
\else
\fi

\usepackage{amsmath}
\usepackage{amssymb}
\usepackage{amsfonts}

\newlength\bshft
\def\fakebold#1{\setbox0=\hbox{$#1$}#1\kern-\wd0\kern\bshft#1\kern-\wd0\kern\bshft#1}

\begin{document}

%
\title{ A Preconditioned Algorithm for Model-Based Iterative CT Reconstruction and Material Decomposition from Spectral CT Data}
%
%
%

\author{Matthew~Tivnan,
        Wenying~Wang,
        and~J.~Webster~Stayman\vspace{-8mm}
}

\maketitle

\begin{abstract}
Model-based material decomposition is a statistical iterative reconstruction framework where basis material density images are estimated directly from spectral CT data. This method uses a physical model for polyenergetic x-ray transmission and attenuation and therefore it does not typically suffer from beam-hardening artifacts. However, this estimation is a poorly-conditioned inverse problem due to the strong anticorrelation between basis materials. In this work we propose an preconditioned optimization algorithm for a nonlinear penalized weighted least-squares objective function.
\end{abstract}


%
\IEEEpeerreviewmaketitle

\vspace{-5mm}
\section{Introduction}
\vspace{-1mm}
Model-based iterative reconstruction (MBIR) was proposed for x-ray computed tomography (CT) over 20 years ago  \cite{erdogan1999ordered} \cite{erdogan1999monotonic}, yet they are only recently starting to be implemented on commercial CT systems due, in large part, to the computational cost \cite{pan2009commercial}. These methods are generally based on \textit{maximum a posteriori} (MAP) estimation or an approximation thereof (e.g. penalized likelihood). Through a combination of advanced physical models, statistical weighting of the data, and the incorporation of prior knowledge via regularization terms, MBIR has consistently demonstrated the potential for improved image quality for the same radiation dose, or the same image quality for a reduced radiation dose. However, to achieve these performance improvements, the algorithm generally involves running many iterations of an image update routine to numerically optimize the MAP objective. 

One such optimization algorithm is the separable parabolic surrogates (SPS) algorithm, originally proposed by \cite{erdogan1999ordered}. It was recently extended an advanced physical and statistical model for the realistic non-idealities of data acquisition with modern multi-slice CT or cone-beam CT systems \cite{tilley2017penalized}. By incorporating effects such as system blur (e.g. focal spot, scintillating detectors), noise correlations, and non-linear polyenergetic effects (e.g. beam hardening) into the reconstruction model, MBIR can result in an inversion of some of these effects (e.g. focal spot deconvolution) to reconstruct high-fidelity images. 

This model can also be extended to material density quantization. If the polyenergetic attenuation from the object is parameterized by a finite number of basis material density images, and the projection measurements have varied sensitivity spectra, as in spectral CT systems, those density images can be estimated in a one-step model-based material decomposition (MBMD) algorithm \cite{tilley2018general} \cite{tilley2018model}.

Due to the large size of the data from multi-slice or cone-beam CT scanners, the various complex geometric and physical models used by MBIR, and the large number of iterations necessary to optimize the MAP objective function, the computational cost of these algorithms is high. For this reason, it is an ongoing area of interest to investigate methods to accelerate the optimization routine. Gradient-based iterative optimization algorithms typically converge slowly for large-scale or poorly-conditioned inverse problems such as the MAP estimator used in MBIR. For that reason, preconditioning techniques can have a dramatically positive impact on convergence. 

The simplest preconditioner is a diagonal scaling matrix. Several algorithms including Expectation Maximization (EM) \cite{lange1984reconstruction}, Scaled Gradient Descent (SGD), \cite{erdogan1999ordered}, and SPS itself can be viewed as diagonal preconditioners applied to the standard gradient descent algorithm. While these diagonal preconditioners are effective and reliable, they are considered to be relatively conservative approximations of the ideal preconditioner which would be the inverse of the hessian of the objective function. 

Non-diagonal preconditioners based on a Fourier basis have have also been applied to a conjugate-gradient-based reconstruction algorithm assuming a linear model. These include Fourier preconditioners which are closely connected to the ramp filter \cite{clinthorne1993preconditioning} and an extension to shift-variant systems using channelized frequency bands \cite{fu2013space}. For MBMD, the conditioning is even worse than conventional CT image reconstruction with MBIR due to the strong correlations between different estimated basis materials. A block-diagonal preconditioner for the primal-dual optimization algorithm was proposed by \cite{sidky2018three}. This strategy dramatically improves the conditioning of the material decomposition portion of the problem.

In this work we propose a general preconditioned version of the SPS algorithm. We define the conditions for the optimal preconditioner and describe a process for finding approximations thereof. We also apply this general preconditioned SPS algorithm to three specific cases: conventional image reconstruction from single-energy CT data, projection-domain decomposition from spectral CT data, and one-step basis material density estimation from spectral CT data in a full preconditioned MBMD algorithm. For the single-energy CT case, we derive a shift-invariant preconditioner based on a fourier basis approximation of the hessian, as well as a shift-variant version based on a haar-wavelet basis approximation. For the projection-domain decomposition case, we describe how to same model can be used to estimate material line integrals (without spatial reconstruction) and we propose a cross-material preconditioner. Finally, for the one-step MBMD case, we propose a block-diagonal cross-material preconditioner, and a cross-material cross-voxel preconditioner based on the fourier basis for each material, as well as a shift-variant version using a haar-wavelet basis.

\section{Methods}

\subsection{Generalized Models for Data Acquisition and Estimation}

For a generalized physical model for x-ray transmission data acquisition, we assume that the projection data, $\mathbf{y}$, are normally distributed with known covariance, $\mathbf{\Sigma_y}$, and mean, $\mathbf{\bar{y}}(\mathbf{x})$ given by \eqref{eq:meanModel}.
    
\begin{equation}
    \mathbf{\bar{y}}(\mathbf{x}) = \fakebold{\mathbb{B}} \exp{( - \fakebold{\mathbb{A}} \mathbf{x}})
    \label{eq:meanModel}
\end{equation}

\noindent where, $\mathbf{x}$, is an unknown parameter vector, and, $\fakebold{\mathbb{A}}$ and $\fakebold{\mathbb{B}}$ are placeholders for linear operators inside and outside the exponential operator, respectively. They are used to  define the expected non-linear relationship between $\mathbf{x}$ and the projection measurements, $\mathbf{y}$. This general formulation can be used to model physical phenomena such as system blur or beam hardening for single-energy CT systems. We will show that, based on the definition of $\mathbf{x}$, this model can also be used for projection-domain decomposition or direct estimation of material density distributions via model-based material decomposition from spectral CT data.

A \emph{maximum a posteriori} estimator for this statistical model can be approximated in terms of the negative log-likelihood of the data $\mathbf{\mathcal{L}}(\mathbf{y}|\mathbf{x})$ and an approximation of the negative log-prior probability of $\mathbf{x}$, given by the penalty $\mathbf{\mathcal{R}}(\mathbf{x})$ as shown in the penalized likelihood objective function shown in \eqref{eq:mapEstimator}.

\begin{gather}
    \mathbf{\hat{x}}{=}\underset{\mathbf{x}}{\text{argmin}}\hspace{2pt}\Phi(\mathbf{x}|\mathbf{y}){=}\underset{\mathbf{x}}{\text{argmin}}\hspace{2pt}\boldsymbol{\mathcal{L}}(\mathbf{y}|\mathbf{x}){+}\boldsymbol{\mathcal{R}}(\mathbf{x}) \label{eq:mapEstimator} \\ 
    \boldsymbol{\mathcal{L}}(\mathbf{y}|\mathbf{x}){=}\frac{1}{2}(\mathbf{y}{-}\mathbf{\bar{y}}(\mathbf{x}))^T \boldsymbol{\Sigma_y}^{-1}(\mathbf{y}{-}\mathbf{\bar{y}}(\mathbf{x}))  
\end{gather}

\noindent For the purposes of this work we will assume the penalty function takes the quadratic form, $\boldsymbol{\mathcal{R}}(\mathbf{x}) = \frac{1}{2}\mathbf{x}^T\mathbf{R}\mathbf{x}$.

The objective function, $\Phi(\mathbf{x})$, is globally convex, but it has no closed-form analytical solution for the global minimum. Instead, estimate $\mathbf{\hat{x}}$ with an iterative optimization algorithm.

\subsection{The Separable Parabolic Surrogates Algorithm}

The concept of operation for the SPS algorithm involves defining a surrogate objective function, $\Tilde{\Phi}(\mathbf{x};\mathbf{x}^{(n)})$ which is separable (i.e. diagonal Hessian) with respect to the elements of $\mathbf{x}$, and majorizes the original objective, $\Phi(\mathbf{x})$. Therefore, we can iteratively update $\mathbf{x}^{(n+1)}$ as the minimum of the surrogate function for which a closed-form solution exists. The full algorithm is given by the following set of formulas.

\begin{gather}
    \Tilde{\Phi}(\mathbf{x};\mathbf{x}^{(n)}) = \boldsymbol{\Tilde{\mathcal{L}}}(\mathbf{x};\mathbf{x}^{(n)}) + \boldsymbol{\Tilde{\mathcal{R}}}(\mathbf{x}) \\
    \frac{\boldsymbol{\partial}\mathbf{\Tilde{L}}}{\boldsymbol{\partial}\mathbf{x}}^{(n)} = \fakebold{\mathbb{A}}^T\Big(D\{\boldsymbol{\gamma}\}\exp{(-2\fakebold{\mathbb{A}}\mathbf{x})} - D\{\boldsymbol{\rho}^{(n)}\}\Big)\\
    \frac{\boldsymbol{\partial^2}\mathbf{\Tilde{L}}}{\boldsymbol{\partial}\mathbf{x}^2}^{(n)} = D\{\fakebold{\mathbb{A}}^TD\{\boldsymbol{\gamma}\}\mathbf{c}^{(n)}\}\\
    \mathbf{x}^{(n+1)} = \mathbf{x}^{(n)} - \Big[\frac{\boldsymbol{\partial^2}\mathbf{\Tilde{L}}}{\boldsymbol{\partial}\mathbf{x}^2}^{(n)} + \frac{\boldsymbol{\partial^2}\mathbf{\Tilde{R}}}{\boldsymbol{\partial}\mathbf{x}^2}\Big]^{-1}\Big[\frac{\boldsymbol{\partial}\mathbf{\Tilde{L}}}{\boldsymbol{\partial}\mathbf{x}}^{(n)} + \frac{\boldsymbol{\partial}\mathbf{\Tilde{R}}}{\boldsymbol{\partial}\mathbf{x}}\Big]
\end{gather}

\noindent where $\boldsymbol{\Tilde{\mathcal{L}}}(\mathbf{x};\mathbf{x}^{(n)})$ and $\boldsymbol{\Tilde{\mathcal{R}}}(\mathbf{x})$ are separable surrogates for the data likelihood and regularization terms of the objective function, respectively. The quantities $\boldsymbol{\eta}$, $\boldsymbol{\gamma}$, $\boldsymbol{\rho}^{(n)}$, and $\mathbf{c}^{(n)}$ are defined below.

\begin{gather}
    \boldsymbol{\eta} = \fakebold{\mathbb{B}}^T \mathbf{\Sigma_y}^{-1}\fakebold{\mathbb{B}}\mathbf{1} \\
    \boldsymbol{\gamma} = \fakebold{\mathbb{A}}\mathbf{1} \\
    \boldsymbol{\rho}^{(n)} = \fakebold{\mathbb{B}}^T \mathbf{\Sigma_y}^{-1}\fakebold{\mathbb{B}} \mathbf{x}^{(n)} - D\{\boldsymbol{\eta}\}\mathbf{x}^{(n)} - \fakebold{\mathbb{B}}^T \mathbf{\Sigma_y}^{-1} \mathbf{y} \\
    \mathbf{c}^{(n)} = f_c\Big(\fakebold{\mathbb{A}}\mathbf{x}^{(n)}\Big)
\end{gather}

\noindent where $f_c(\mathbf{x}^{(n)})$ is the maximum curvature calculation described in \cite{tilley2017penalized}. 

While this algorithm guarantees that the cost will monotonically decrease with each iteration, the separable approximation of the objective function will be very poor if the estimates, $\mathbf{x}$, are strongly inter-correlated, leading to a small step size and slow convergence. Therefore, we propose to derive a preconditioned version of the algorithm to find a new basis for estimation for which separability is a more effective approximation.

\subsection{Preconditioned Optimization}

The Hessian of the objective function evaluated at the solution, $\mathbf{\hat{x}}$, is given by $\mathbf{H} = \frac{\boldsymbol{\partial^2}\Phi}{\boldsymbol{\partial}\mathbf{x}^2}(\mathbf{\hat{x}}) $. For the penalized likelihood objective in \eqref{eq:mapEstimator}, this hessian is

\begin{gather}
    \mathbf{H} = \fakebold{\mathbb{A}}^T\mathbf{D}^T\fakebold{\mathbb{B}}^T \mathbf{\Sigma}_\mathbf{y}^{-1} \fakebold{\mathbb{B}}\mathbf{D}\fakebold{\mathbb{A}} + \mathbf{R} \\
    \mathbf{D} = D\{\exp{(-\fakebold{\mathbb{A}} \mathbf{\hat{x}})}\}
\end{gather}

\noindent Note that for many cases, an effective approximation for $\mathbf{D}$ is available without dependence on $\mathbf{\hat{x}}$.

In many cases, there are strong cross-estimate correlations that make this a poorly-conditioned inverse problem to estimate $\mathbf{\hat{x}}$ given $\mathbf{y}$. We propose to find a linear transformation to a new basis, $\mathbf{x'}$, which improves the conditioning of the estimation problem. This transformation is defined by the relationship, $\mathbf{x}~=~\mathbf{M}~\mathbf{x'}$. Therefore, the new hessian, $\mathbf{H'}$, associated with the preconditioned estimation of $\mathbf{\hat{x'}}$ is 

\begin{gather}
    \mathbf{H'} = \frac{\boldsymbol{\partial^2}\Phi}{\boldsymbol{\partial}\mathbf{x'}^2}~(\mathbf{\hat{x'}}) = \Big(\frac{\boldsymbol{\partial}\mathbf{x}}{\boldsymbol{\partial}\mathbf{x'}}\Big)^T \frac{\boldsymbol{\partial^2}\Phi}{\boldsymbol{\partial}\mathbf{x}^2} \Big(\frac{\boldsymbol{\partial}\mathbf{x}}{\boldsymbol{\partial}\mathbf{x'}}\Big) = \mathbf{M}^T\mathbf{H}\mathbf{M}
\end{gather}

Therefore, effective preconditioner will satisfy, $\mathbf{M}^T\mathbf{H}\mathbf{M} \approx \mathbf{I}$, or at least the condition number of $\mathbf{M}^T\mathbf{H}\mathbf{M}$ will be much less than that of $\mathbf{H}$. 

Substituting  $\mathbf{M}\mathbf{x'}$ into \eqref{eq:meanModel} gives

\begin{equation}
    \mathbf{y} = \fakebold{\mathbb{B}} \exp{(-\fakebold{\mathbb{A}} \mathbf{M} \mathbf{x'})}
\end{equation}

\noindent Based on this form we propose to use a preconditioned version of the SPS algorithm. This involves replacing the linear operator, $\fakebold{\mathbb{A}}$, with the preconditioned version, $\fakebold{\mathbb{A}} \mathbf{M}$, estimating $\mathbf{x'}$ via the SPS algorithm, and finally applying $\mathbf{x}~=~\mathbf{M}~\mathbf{x'}$ to reproject the estimates to the original basis.

Since $\mathbf{H}$ is symmetric by construction and assumed to be full-rank, we will aim to find a precondtitioner which approximates the unique symmetric matrix square root of the inverse Hessian (i.e. $\mathbf{M} = \mathbf{H}^{-\frac{1}{2}}$). The eigen decomposition of $\mathbf{H}$ is given by

\begin{equation}
    \mathbf{H} = \mathbf{U_H}^T \mathbf{\Lambda_H} \mathbf{U_H}
    \label{eq:hessianEigenBasis}
\end{equation}

\noindent where $\mathbf{U_H}$ is an orthogonal matrix containing the eigenvectors of the hessian, $\mathbf{H}$, and $\mathbf{\Lambda_H}$ is a diagonal matrix containing the eigenvalues. Therefore, the ideal preconditioner is given by

\begin{equation}
    \mathbf{M} = \mathbf{U_H}^T \mathbf{\Lambda}_\mathbf{H}^{-\frac{1}{2}} \mathbf{U_H} 
    \label{eq:preconditionerEigenBasis}
\end{equation}

In the following sections we will propose preconditioner designs for a few specific applications by finding an orthogonal basis $\mathbf{\Tilde{U}_H}$ which is anapproximation of the eigenbasis of the hessian. Then we will find the corresponding approximate eigenvalues, $\mathbf{\Tilde{\Lambda}_H}$ according to the following formula.

\begin{equation}
    \mathbf{\Tilde{\Lambda}_H} = D\{\mathbf{\Tilde{U}_H} \mathbf{ \Delta x}_\text{test}\}^{-1} D\{\mathbf{\Tilde{U}_H} \mathbf{H} \boldsymbol{\delta x}_\text{test}\}
    \label{eq:eigenApproximation}
\end{equation}

\noindent which will result in an approximation of the hessian given by $\mathbf{\Tilde{H}}=\mathbf{\Tilde{U}_H}^T \mathbf{\Tilde{\Lambda}_H}\mathbf{\Tilde{U}_H}$ which matches the response of $\mathbf{H}$ for some test function, $\boldsymbol{\delta x}_\text{test}$. Note that $\boldsymbol{\delta x}_\text{test}$ should be designed appropriately to excite all of the eigenvectors such that $D\{\mathbf{\Tilde{U}_H} \mathbf{ \Delta x}_\text{test}\}]^{-1}$ is defined. Finally, we can substitute the approximations $\mathbf{\Tilde{U}_H}$ and $\mathbf{\Tilde{\Lambda}_H}$  into \eqref{eq:preconditionerEigenBasis} to establish a establish a preconditioned optimization algorithm.

\subsection{Standard CT Reconstruction}

For standard CT reconstruction, we aim to estimate an image composed of attenuation coefficients, $\boldsymbol{\mu}$, given standard single-energy CT data, $\mathbf{y}$. Therefore, we have the following definitions.

\begin{gather}
    \mathbf{x} \xrightarrow{} \boldsymbol{\mu} \\
    \fakebold{\mathbb{A}} \xrightarrow{} \mathbf{A} \\
    \fakebold{\mathbb{B}} \xrightarrow{} \mathbf{G} \mathbf{B} \\
    \mathbf{\bar{y}}(\boldsymbol{\mu}) = \mathbf{G} \mathbf{B} \exp\{ - \mathbf{A} \boldsymbol{\mu} \}
\end{gather}

\noindent where $\mathbf{A}$ is the forward projector, capturing the system's spatial sampling geometry, $\mathbf{B}$ models the projection-domain blur (e.g. focal spot, scintilating detectors), and $\mathbf{G}$ is the gain. Note that this particular model does not include a polyenergetic absorption model and would therefore be subject to beam-hardening artifacts if applied to measurements acquired with a polyenergetic source. This leads to the following definition of the Hessian.

\begin{gather}
    \mathbf{H} = \frac{\boldsymbol{\partial^2}\Phi}{\boldsymbol{\partial}\boldsymbol{\mu}^2}(\boldsymbol{\hat{\mu}}) = \mathbf{F} + \mathbf{R} \\
    \mathbf{F} =  \mathbf{A}^T \mathbf{D}^T \mathbf{B}^T \mathbf{G}^T \boldsymbol{\Sigma_y}^{-1} \mathbf{G} \mathbf{B}  \mathbf{D} \mathbf{A}\\
    \mathbf{D} = D\{\exp{({-}\mathbf{A}\boldsymbol{\hat{\mu}})}\}
\end{gather}

An effective approximation for $\mathbf{D}$ is available directly from the gain-corrected measured data. The vector $\boldsymbol{\mu}$ is sized $N_j \times 1$ and therefore the Fisher information matrix $\mathbf{F}$ and the quadratic regularization matrix $\mathbf{R}$ are sized $N_j \times N_j$. The following is a quadratic smoothness regularizer, for example.

\begin{gather}
    R_{j,j'} = 
    \begin{array}{cc}
  \Big\{ & 
    \begin{array}{cc}
      |\mathcal{N}_j| & j = j' \\
      -1 &  j \in \mathcal{N}_j \\
      0 & \text{otherwise}
    \end{array}
    \end{array}
    \label{eq:quadraticSmoothess}
\end{gather}

\noindent where $\mathcal{N}_j$ is a neighborhood around voxel $j$ (not including $j$) and $|\mathcal{N}_j|$ is the cardinality of that neighborhood.

\subsubsection{Fourier Preconditioner} $\enspace$ 

To begin the derivation of a cross-voxel preconditioner, consider the case where $\mathbf{A}$ represents the discrete radon transform, corresponding to a parallel beam imaging geometry, and $\mathbf{D}$, $\mathbf{B}$, $\mathbf{G}$, and $\mathbf{\Sigma_y}$ are all diagonal. For that special case, $\mathbf{F}^{-1} = (\mathbf{A}^T\mathbf{A})^{-1}$ is the ramp filter used in filtered back-projection, so $\mathbf{F}$ takes the form shown below.

\begin{equation}
    \mathbf{F} = \mathbf{U_{DFT}}^T \mathbf{\Lambda}_F \mathbf{U_{DFT}}
\end{equation}

\noindent where $\mathbf{U_{DFT}}$ is the unitary discrete fourier transform, which is the eigenbasis of $\mathbf{F}$, and $\mathbf{\Lambda_F}$ is diagonal and contains the eigenvalues, or the frequency coefficients for the inverse ramp filter. This form implies that $\mathbf{F}$ is a shift-invariant operator. For the example in \eqref{eq:quadraticSmoothess}, the matrix operator $\mathbf{R}$ is also shift-invariant and can therefore also be diagonalized by the discrete fourier transform. Therefore, the eigenbasis of $\mathbf{H}$ is also the a fourier basis or $\mathbf{U_H} = \mathbf{U_{DFT}}$, and we can solve for $\mathbf{\Lambda_H}$ exactly using the formula by applying \eqref{eq:eigenApproximation} as shown below.

\begin{equation}
    \mathbf{\Lambda_H} = \mathbf{\Tilde{\Lambda}_H} = D\{\mathbf{U_{DFT}} \boldsymbol{\delta_j}\}^{-1} D\{\mathbf{U_{DFT}} [\mathbf{F} + \mathbf{R}] \boldsymbol{\delta_j}\}
    \label{eq:fourierEigenbasis}
\end{equation}

\noindent where $\boldsymbol{\delta_j}$ is an impulse with a small value only at voxel~$j$.

Although we understand that for the general case where $\mathbf{A}$ may represent a non-parallel geometry or for non-diagonal matrices $\mathbf{D}$, $\mathbf{B}$, $\mathbf{G}$, and $\mathbf{\Sigma_y}$, the response of $\mathbf{F}$ will be shift-invariant, we can still make the shift-invariant approximation $\mathbf{U_H} \approx \mathbf{U_{DFT}}$ and apply \eqref{eq:fourierEigenbasis} to establish an approximate set of eigenvalues (or filter coefficients) for a shift-invariant preconditioner. The result will be an approximate model for the hessian that perfectly matches the frequency response of $\mathbf{H}$ at position $j$, but it may be a poor approximation for positions $j'$ which are far from $j$. This Fourier preconditioner is therefore constructed as follows.

\begin{equation}
    \mathbf{M} = \mathbf{U_{DFT}}^T \mathbf{\Tilde{\Lambda}_H}^{-\frac{1}{2}} \mathbf{U_{DFT}}
\end{equation}

This is a desirable preconditioner because it is essentially a linear filter which can be applied using fast fourier transform algorithm, a diagonal scaling matrix, and the inverse fourier transform. This is very similar to the preconditioner proposed in \cite{clinthorne1993preconditioning}. The pitfall of this approach is that for cases where the operator $\mathbf{H}$ is strongly shift-invariant, this approximation of the hessian may be very poor.

\subsubsection{Modified Fourier Preconditioner} $\enspace$

The Fourier preconditioner described above assumes the same frequency response at every position in the image domain. One straight-forward modification to this algorithm is to first normalize the zero-frequency response before assuming shift invariance. This modification extends the model to an operator with relative frequency weights which are shift invariant with a shift-variant overall scale. 

The zero-frequency response normalization is given by the following diagonal preconditioner.

\begin{equation}
    \mathbf{M_0} = D\{[\mathbf{F} + \mathbf{R}]\mathbf{1}\}^{-\frac{1}{2}}
    \label{eq:diagonalPreconditioner}
\end{equation}

\noindent Then, we assume the normalized hessian is a Fourier system. 

\begin{equation}
    \mathbf{M_0}^T \mathbf{H} \mathbf{M_0} = \mathbf{U_{DFT}}^T \mathbf{\Tilde{\Lambda}_H} \mathbf{U_{DFT}}
\end{equation}

\noindent and solve for the eigenvalues accordingly.

\begin{equation}
    \mathbf{\Tilde{\Lambda}_H} = D\{\mathbf{U_{DFT}} \boldsymbol{\delta_j}\}^{-1} D\{\mathbf{U_{DFT}}  \mathbf{M_0}^T[\mathbf{F} + \mathbf{R}] \mathbf{M_0} \boldsymbol{\delta_j}\}
\end{equation}

The modified Fourier preconditioner is therefore given by the following formula.

\begin{equation}
    \mathbf{M} = \mathbf{M_0}^T \mathbf{U_{DFT}}^T \mathbf{\Tilde{\Lambda}_H}^{-\frac{1}{2}} \mathbf{U_{DFT}} \mathbf{M_0}
\end{equation}

This combination of a diagonal preconditioner and a Fourier preconditioner is capable of approximating the shift-variant scale but still assumes a shift-invariant relative frequency response. 

\subsubsection{Wavelet Precondtioner} $\enspace$ 

For a shift-variant preconditioner, we propose to approximate the eigenbasis of $\mathbf{H}$ with a unitary discrete wavelet transform, that is, $\mathbf{U_H} \approx \mathbf{U_{DWT}}$. Examples of unitary wavelet transforms include the Haar transform and other unitary transforms based on Daubechies wavelets \cite{daubechies1992ten}.  

We have chosen this particular basis because wavelets are capable of encoding information about both frequency response and shift-variant behaviors. 

To find the eigenvalues, we use a test function $\boldsymbol{\delta x}_\text{test}$~$=$~$\mathbf{U_{DWT}}^T \mathbf{1}$ which is excites all eigenvectors of the wavelet basis . This is analogous to the impulse function, $\boldsymbol{\delta_1}$~$=$~$\mathbf{U_{DFT}}^T \mathbf{1}$, which excites all eigenvectors of the Fourier basis. 

Following a similar procedure as described in the previous section, the zero-frequency response is normalized via the diagonal preconditioner shown in \eqref{eq:diagonalPreconditioner}. Then, assuming a wavelet eigenbasis, the approximate eigenvalues of the normalized hessian, $\mathbf{M_0}^T \mathbf{\Tilde{\Lambda}_H} \mathbf{M_0}$, can be established using \eqref{eq:eigenApproximation} and the preconditioner can be formulated as

\begin{equation}
        \mathbf{M} = \mathbf{M_0}^T \mathbf{U_{DWT}}^T \mathbf{\Tilde{\Lambda}_H}^{-\frac{1}{2}} \mathbf{U_{DWT}} \mathbf{M_0}
\end{equation}

In general, the pitfalls of the wavelet transform are that there are relatively few number of eigenvectors used to describe the shift-variance of the low-frequency response, and there is also an upper limit on frequencies that are captured. 

This combination of a diagonal preconditioner and a wavelet preconditioner is capable of modeling a shift-variant local frequency response.

\subsection{Projection-Domain Decomposition}

For model-based projection-domain material decomposition, we aim to estimate material line integrals, $\boldsymbol {\ell}$, given spectral CT data, $\mathbf{y}$ which is composed of projections with varied spectral sensitivity arranged into channels. Typically there are at least as many channels as materials. Therefore, we have the following definitions.

\begin{gather}
    \mathbf{x} \xrightarrow{} \boldsymbol {\ell} \\
    \fakebold{\mathbb{A}} \xrightarrow{} \mathbf{Q} \\
    \fakebold{\mathbb{B}} \xrightarrow{}  \mathbf{G} \mathbf{B} \mathbf{S} \\
    \mathbf{\bar{y}}(\boldsymbol{\ell}) = \mathbf{G} \mathbf{B} \mathbf{S} \exp\{ - \mathbf{Q} \boldsymbol{\ell} \}
\end{gather}

\noindent where $\mathbf{Q}$ maps material basis-material line integrals to projection-domain attenuation spectra using a weighted sum of the attenuation spectra for each basis material, $\mathbf{S}$ is the projection-dependent system spectral sensitivity for each channel of the spectral CT system, $\mathbf{B}$ models the channel-dependent projection-domain blur, and $\mathbf{G}$ is the gain. This leads to the following definition of the Hessian.

\begin{gather}
    \mathbf{H} = \frac{\boldsymbol{\partial^2}\Phi}{\boldsymbol{\partial}\boldsymbol{\ell}^2}(\boldsymbol{\hat{\ell}}) = \mathbf{F} + \mathbf{R} \\
    \mathbf{F} =  \mathbf{Q}^T \mathbf{D}^T \mathbf{S}^T \mathbf{B}^T \mathbf{G}^T \boldsymbol{\Sigma_y}^{-1} \mathbf{G} \mathbf{B}  \mathbf{S} \mathbf{D} \mathbf{Q}\\
    \mathbf{D} = D\{\exp{({-}\mathbf{Q}\boldsymbol{\hat{\ell}})}\}
\end{gather}

An effective approximation for $\mathbf{D}$ can be obtained by assuming all of the attenuation was due to water and conducting an inexpensive single-material line integral estimation. The vector $\boldsymbol{\ell}$ is sized $N_i N_k \times 1$, so the hessian is sized $N_i N_k \times N_i N_k$. Here $i$ is used to index projections and $k$ is used to index materials.  

\subsubsection{Cross-Material Preconditioner} $\enspace$

Since the estimation of $\boldsymbol{\ell}$ occurs in the projection-domain,  different positions can be effectively approximated as separable. That is, the elements of the cross-projection second-derivatives in the hessian can be approximated as zero, or $H_{ik,i'k} \approx 0 \enspace$ for $\enspace i \neq i'$. Therefore, an approximate preconditioner can be constructed from $N_k \times N_k$ blocks, denoted as $\mathbf{\Tilde{H}_i}$ for each projection. There are $N_i N_k$ values which compose column $k$ of those blocks, denoted as the $N_iN_k \times 1$ vector $\mathbf{\Tilde{H}_k}$, which can be computed as

\begin{equation}
    \mathbf{\Tilde{H}_k} = [\mathbf{F} + \mathbf{R}] \mathbf{1_k} 
    \label{eq:materialColumns}
\end{equation}

\noindent where the vector $\mathbf{1_k}$ has a flat response of ones for material $k$ and zeros for other materials. 

After $N_k$ applications of \eqref{eq:materialColumns}, the block-diagonal hessian can be reorganized into the blocks, $\mathbf{\Tilde{H}_i}$, and the preconditioner for can be constructed as  $\mathbf{M_i} = \mathbf{\Tilde{H}_i}^{-\frac{1}{2}}$. This leads to a block-diagonal preconditioner $\mathbf{M}$ which assumes separability across projections, but applies a non-diagonal preconditioning accross materials.

\subsection{One-Step Reconstruction and Material Decomposition}

In direct model-based material decomposition, we aim to estimate material density maps, $\boldsymbol{\rho}$, given spectral CT data, $\mathbf{y}$ which is composed of projections with varied spectral sensitivity. Therefore, we have the following definitions.

\begin{gather}
    \mathbf{x} \xrightarrow{} \boldsymbol{\rho} \\
    \fakebold{\mathbb{A}} \xrightarrow{} \mathbf{Q} \mathbf{A} \\
    \fakebold{\mathbb{B}} \xrightarrow{} \mathbf{G} \mathbf{B} \mathbf{S}
\end{gather}

\noindent where $\mathbf{A}$ is the forward projector, capturing the system's spatial sampling geometry, $\mathbf{Q}$ contains the mass attenuation spectra for each basis material, $\mathbf{S}$ is the projection-dependent system spectral sensitivity, $\mathbf{B}$ models the projection-domain blur (e.g. focal spot, scintilating detectors), and $\mathbf{G}$ is the gain. This leads to the following definition of the Hessian.

\begin{gather}
    \mathbf{H} = \frac{\boldsymbol{\partial^2}\Phi}{\boldsymbol{\partial}\boldsymbol{\rho}^2}(\boldsymbol{\hat{\rho}}) = \mathbf{F} + \mathbf{R} \\
    \mathbf{F} =  \mathbf{A}^T \mathbf{Q}^T \mathbf{D}^T \mathbf{S}^T \mathbf{B}^T \mathbf{G}^T \boldsymbol{\Sigma_y}^{-1} \mathbf{G} \mathbf{B} \mathbf{S} \mathbf{D} \mathbf{Q} \mathbf{A}\\
    \mathbf{D} = D\{\exp{({-}\mathbf{Q}\mathbf{A}\boldsymbol{\hat{\rho}})}\}
\end{gather}

\noindent We assume that $\mathbf{R}$ is the hessian of a cross-material quadratic smoothness penalty, $\boldsymbol{\mathcal{R}}(\boldsymbol{\rho})$, which we define as

\begin{equation}
    \boldsymbol{\mathcal{R}}(\boldsymbol{\rho}){=}\frac{1}{2}\sum_{k = 1}^{N_k} \sum_{k' = 1}^{N_k} \beta_{k,k'}\sum_{j = 1}^{N_j}\sum_{j' \in \mathcal{N}_j} (\rho_{jk}{-}\rho_{j'k})(\rho_{jk'}{-}\rho_{j'k'})
    \label{quadReg_summation}
\end{equation}

\noindent where $\rho_{jk}$ is the denisty of basis material $k$ at voxel  $j$, the set $\mathcal{N}_j$ is the set of voxels neighboring $j$ (not including $j$) which has cardinality $|\mathcal{N}_j|$, and the parameters, $\beta_{k,k'} > 0$ are the cross-material regularization weights. We can write $\mathbf{R}$ in terms of the two symmetric matrices $\boldsymbol{\beta}$, modeling the regularization strength, and $\mathbf{\Gamma}$ modeling the kernel for the spatial smoothness penalty as follows.

\begin{gather}
    \mathbf{R} = \boldsymbol{\beta}^{1/2}\mathbf{\Gamma}\boldsymbol{\beta}^{1/2} = \mathbf{\Gamma}^{1/2}\boldsymbol{\beta}\mathbf{\Gamma}^{1/2} \\
    \beta_{jk,j'k'} = 
    \begin{array}{cc}
  \Big\{ & 
    \begin{array}{cc}
      \beta_{k,k'} & j = j' \\
      0 & \text{otherwise}
    \end{array}
\end{array} \\
    \Gamma_{jk,j'k'} = 
    \begin{array}{cc}
  \Big\{ & 
    \begin{array}{cc}
      |\mathcal{N}_j| & k = k' , j = j' \\
      -1 & k = k' , j \in \mathcal{N}_j \\
      0 & \text{otherwise}
    \end{array}
    \end{array} 
\end{gather}

\subsubsection{Cross-Material Precondtioner} $\enspace$ 

The correlations between estimates for different materials severely impacts the the conditioning of the estimation problem for model-based material decomposition. Therefore we will define $\mathbf{\Tilde{H}}$ as shown below in order to define a shift-variant cross-material preconditioner.

\begin{equation}
    \Tilde{H}_{jk,j'k'} = 
  \Big\{ 
    \begin{array}{cc}
       H_{jk,j'k'}& j = j' \\
      0 & \text{otherwise}
    \end{array}
\end{equation}

Therefore, $\mathbf{H}$ can be constructed from $N_j$ blocks denoted by $\mathbf{\Tilde{H}_j}$ which are sized $N_k \times N_k$. Column $j$ of all of those blocks can be organized into a $N_k N_j \times 1$ vector, denoted by $\mathbf{\Tilde{H}_k}$, which can be computed as

\begin{equation}
    \mathbf{\Tilde{H}_k} = [\mathbf{F} + \mathbf{R}] \mathbf{1_k}
\end{equation}

\noindent where $\mathbf{1_k}$ is an image of ones for material $k$ and zero for the other materials. Then, the hessian can be reorganized into $N_j$ blocks, $\mathbf{\Tilde{H}_j}$. Then, a block-diagonal preconditioner, $\mathbf{M_0}$, which has a block for each voxel equal to $\mathbf{\Tilde{H}_j}^{-\frac{1}{2}}$.

This block-diagonal preconditioner matches the zero-frequency response of the hessian including the cross-material zero-frequency response. However, this format does not handle the cross-voxel correlations in the estimates or the more abstract cross-voxel-cross-material correlations.

\subsubsection{Cross-Material Fourier Precondtioner} $\enspace$ 

In this section we seek to derive a cross-material Fourier preconditioner to handle both cross-material and cross-voxel correlations. We begin with the approximation that after normalizing the zero-frequency response, the modified hessian operator is approximately shift-invariant. Note that this shift-invariance describes both the in-basis response as well as the cross-basis response of the normalized hessian. We express this in the formula below.

\begin{equation}
    \mathbf{M_0}^T \mathbf{H} \mathbf{M_0} \approx \mathbf{U_{DFT}}^T \mathbf{L} \mathbf{U_{DFT}}
\end{equation}

\noindent $\mathbf{U_{DFT}}$ represents the unitary discrete Fourier transform applied individually and identically to all materials. The matrix $\mathbf{L}$ contains weights applied to spatial the $N_j$ spatial frequencies. If all spatial frequencies follow the same cross-material relationship, then the spatial correlations and material correlations would be entirely separable and $\mathbf{L}$ would be a diagonal matrix. In general, this may not be the case so we assume $\mathbf{L}$ is block diagonal contain $N_j$ blocks, one fore each spatial frequency, denoted as $\mathbf{L_j}$, each sized $N_k \times N_k$. We can compute the column $k$ for all $N_j$ blocks organized into the $N_j N_k \times 1$ vector, $\mathbf{L_k}$, as follows.

\begin{equation}
    \mathbf{L_k} = D\{\mathbf{U_{DFT}}\boldsymbol{\delta_{jk}}\}^{-1}\Big[\mathbf{U_{DFT}}\mathbf{M_0}^T [\mathbf{F} + \mathbf{R}] \mathbf{M_0} \boldsymbol{\delta_{jk}}\Big]
    \label{eq:localImpulseResponse}
\end{equation}

\noindent The formula above describes the Fourier transform of the zero-frequency-normalized hessian ($\mathbf{M_0}^T \mathbf{H} \mathbf{M_0}$) applied to $\boldsymbol{\delta_{jk}}$, an impulse in material~$k$ at position~$j$, divided by the frequency response of $\boldsymbol{\delta_{jk}}$ to correct for the phase. Therefore, the vector $\mathbf{L_k}$ represents the local response of all materials and spatial frequencies to an impulse in material~$k$ at position~$j$. 

After rearranging the columns, $\mathbf{L_k}$, into the $N_j$ blocks, $\mathbf{L_j}$, the cross-material Fourier preconditioner can be constructed by taking the symmetric matrix negative square root of each of the blocks as written below. 

\begin{equation}
    \mathbf{M} = \mathbf{M_0}^T \mathbf{U_{DFT}}^T \mathbf{L}^{-\frac{1}{2}} \mathbf{U_{DFT}} \mathbf{M_0}
\end{equation}

This is the cross-material analog of the modified Fourier preconditioner defined in a previous section for standard CT reconstructions. It makes the assumption that after normalization of the zero-frequency response (including cross-material), the modified hessian is approximately shift-invariant.  This form will perfectly match the local impulse response (including cross-material) for at least one position.

\subsubsection{Cross-Material Wavelet Precondtioner} $\enspace$ 

Instead of assuming a Fourier basis for each material as in the previous section, we can assume a wavelet basis of the zero-frequency-normalized hessian as shown below.

\begin{equation}
    \mathbf{M_0}^T \mathbf{H} \mathbf{M_0} \approx \mathbf{U_{DWT}}^T \mathbf{K} \mathbf{U_{DWT}}
\end{equation}

\noindent we can solve for the block-diagonal matrix $\mathbf{K}$ sequentially for each material as formulated below.

\begin{gather}
    \mathbf{K_k} = D\{\mathbf{U_{DWT}}\boldsymbol{\delta_{jk}}\}^{-1}\Big[\mathbf{U_{DWT}}\mathbf{M_0}^T [\mathbf{F} + \mathbf{R}] \mathbf{M_0} \boldsymbol{\delta_{jk}}\Big] \\
    \mathbf{\Delta x} = \mathbf{U_{DWT}}^T \mathbf{1}
\end{gather}

\noindent where $\mathbf{U_{DWT}}$ represents a unitary discrete wavelet transform (such as the Haar transform) applied individually and identically for each material, and the test function $\mathbf{\Delta x}$ is designed to excite all eigenvectors of the wavelet basis just as an impulse function was used to excite all eigenvectors of a Fourier basis.

After rearranging the block-diagonal matrix $\mathbf{L}$, the cross-material wavelet preconditioner can be written as

\begin{equation}
    \mathbf{M} = \mathbf{M_0}^T \mathbf{U_{DWT}}^T \mathbf{K}^{-\frac{1}{2}} \mathbf{U_{DWT}} \mathbf{M_0}
\end{equation}

This form is capable of modeling complicated cross-material and cross-voxel correlations with relatively few total paramters. The cross-material preconditioner $\mathbf{M_0}$ inverts the zero-frequency response of the hessian accross all positions, and the wavelet-domain operator paramterized by the block diagonal matrix $\mathbf{K}$ can be described as a model for the residual shift-variant cross-material spatial frequency response. 

\section{Conclusion}

The slow computational speed of MBIR is a major factor preventing widespread implementation of clinical systems. The SPS algorithm is able to monotonically decrease the objective function in an iterative optimization algorithm. However, if the estimates are highly correlated, the separable surrogate objective function will be a poor match to the truth, leading to very small step sizes. 

Effective preconditioners are an efficient way to improve the conditioning of the inverse problem. In this work, we have proposed a preconditioned version of the SPS algorithm. We have also derived specific preconditioners for applications including standard CT reconstruction, projection-domain material decomposition, and one-step direct model-based material decomposition and image reconstruction. 
Some of the non-diagonal preconditioners presented in this paper (e.g. Fourier preconditioner) have been previously proposed in a similar form. We have also presented novel preconditioning strategies such as the wavelet preconditioner and cross-material preconditioners. 

With effective preconditioning, there is a potential to the computation time necessary for MBIR. These preconditioned optimization algorithms could accelerate reconstructions of standard CT and spectral CT data in a way that makes MBIR more practical for clinical implementation on commercial systems.



\end{document}